\documentclass[aoas,MSNbibl,nameyear,dvips]{arximspdf}
\usepackage{mathbh,dcolumn}
\usepackage{graphicx}
\usepackage{url,breakurl}

\doi{10.1214/14-AOAS743} 
\volume{8}
\issue{3}
\pubyear{2014}
\firstpage{1612}
\lastpage{1639}
\docsubty{FLA}

\makeatletter
\newcolumntype{d}[1]{D{.}{.}{#1}}
\newcommand{\mttt}{\mathrm}
\newcommand{\rrVert}{\Vert}
\newcommand{\llVert}{\Vert}
\newcommand{\pkg}[1]{\textit{#1}}
\newcommand{\NegBin}{\operatorname{NegBin}} 
\newcommand{\sd}{\mathrm{SD}} 
\makeatother

\begin{document}
\begin{frontmatter}

\title{Power-law models for infectious disease spread\thanksref{T1}}
\runtitle{Power-law models for infectious disease spread}

\begin{aug}
\author[A]{\fnms{Sebastian}~\snm{Meyer}\corref{}\ead[label=e1]{Sebastian.Meyer@uzh.ch}\ead[label=u1,url]{www.biostat.uzh.ch}}
\and
\author[B]{\fnms{Leonhard}~\snm{Held}\ead[label=e2]{Leonhard.Held@uzh.ch}}
\runauthor{S. Meyer and L. Held}
\affiliation{University of Zurich}
\address[A]{Epidemiology, Biostatistics and Prevention Institute\\
Department of Biostatistics\\
University of Zurich\\
Hirschengraben 84\\
CH-8001 Z{\"u}rich\\
Switzerland\\
\printead{e1}\\
\phantom{E-mail: }\printead*{e2}\\
\printead{u1}} 
\end{aug}
\thankstext{T1}{Funded by the Swiss National Science Foundation
(project \#137919).}

\received{\smonth{8} \syear{2013}}
\revised{\smonth{2} \syear{2014}}

%
\begin{abstract}
Short-time human travel behaviour can be described by a power law with
respect to distance.
We incorporate this information in space--time models for infectious
disease surveillance data to better capture the dynamics of disease spread.
Two previously established model classes are extended, which both decompose
disease risk additively into endemic and epidemic components:
a~spatio-temporal point process model for individual-level data
and a multivariate time-series model for aggregated count data.
In both frameworks, a power-law decay of spatial interaction is
embedded into the epidemic component and estimated jointly with all
other unknown parameters using (penalised) likelihood inference.
Whereas the power law can be based on \mbox{Euclidean} distance in the point
process model, a novel formulation is proposed for count data where
the power law depends on the order of the neighbourhood of discrete spatial
units.
The performance of the new approach is investigated by a reanalysis of
individual cases of invasive meningococcal disease in Germany
(2002--2008) and count data on influenza in 140 administrative
districts of Southern Germany (2001--2008). In both applications, the
power law substantially improves model fit and predictions, and is
reasonably close to alternative qualitative formulations, where
distance and order of neighbourhood, respectively, are treated as a factor.
Implementation in the \textsf{R}~package \textit{surveillance} allows
the approach to be applied in other settings.
\end{abstract}

%
\end{frontmatter}

\section{Introduction} \label{secintro}

The surveillance of infectious diseases constitutes a key issue of
public health and modelling their spread is basic to the prevention and control
of epidemics.
An important task is the timely detection of disease outbreaks, for which
popular methods are the Farrington algorithm
[\citet{farringtonetal1996,noufailyetal2013}] and cumulative
sum (CUSUM)
likelihood ratio detectors inspired by statistical process control
[\citet{hoehlepaul2008,hoehleetal2009}].
As opposed to such prospective surveillance,
retrospective surveillance 
is concerned with explaining the spread of an epidemic through statistical
modelling, thereby assessing the role of environmental and socio-demographic
factors or contact networks in shaping the evolution of an epidemic.
The spatio-temporal data for such modelling primarily originate from routine
public health surveillance of
the occurrence of infectious diseases and is ideally accompanied by additional
data on influential factors to be accounted for.
Surveillance data are available in different spatio-temporal resolutions,
each type requiring an appropriate model framework.

This paper covers both
a spatio-temporal point process model for indi\-vidual-level data
[proposed by
\citet{meyeretal2011} and motivated by the work of \citet
{hoehle2009}]
and a multivariate time-series model for aggregated count data
[established by \citet{heldpaul2012} and earlier work].
Although these two models are designed for different types of spatio-temporal
surveillance data, both are inspired by the approach of \citet
{heldetal2005}
decomposing 
disease risk additively into ``endemic'' and ``epidemic'' components.
The endemic component captures
exogenous factors such as population, socio-demographic variables, long-term
trends, seasonality, climate, or concurrent incidence of related diseases
(all varying in time and/or space).
Explicit dependence between cases, that is, infectiousness, is then
introduced through epidemic components driven by the observed past.

To describe disease spread in space,
both models account for spatial interaction between units or individuals,
respectively, but up to now, this has been incorporated rather
crudely.
The point process model used a Gaussian kernel to capture spatial interaction,
and the multivariate time-series model restricted epidemic spread from
time $t$
to $t+1$ to adjacent regions.
However, a simple form of dispersal
can be motivated by the findings of \citet{brockmannetal2006}: they
inferred from the dispersal of bank notes in the United States that (short-time)
human travel behaviour can be well described by a decreasing power law
of the distance $x$, that is, $f(x) \propto x^{-d}$ with positive decay
parameter $d$. An important characteristic of this power law is its slow
convergence to zero (``heavy tail''), which in our application enables
occasional long-range transmissions of infectious agents in addition to
principal short-range infections.
In the words of \citet{brockmannetal2006}, their results
``can serve as a starting point for the development of a new class of
models for the spread of human infectious diseases''.
Power laws are well known from the work by \citet{Pareto1896} for the
distribution of income and \citet{Zipf1949} for city sizes and
word frequencies
in texts.
They describe the distribution of earthquake magnitudes
[\citet{gutenbergrichter1944}] and many other natural phenomena
[see \citet{newman2005,pintoetal2012}, for a review of power laws].
\citet{liljerosetal2001} reported on a power-law distribution of
the number of
sexual partners, and \citet{albertbarabasi2002} review recent advances
in network theory including scale-free networks where the number of
edges is distributed according to a power law.
Interestingly, a power law was also used as the distance decay function in
geographic profiling for serial violent crime investigation
[\citet{Rossmo2000}]
as well as in an application of this technique to 
identify environmental sources of infection
[\citet{lecomberetal2011}]. 
Examples of power-law transmission kernels to model the spatial dynamics
of infectious diseases can be found in plant epidemiology
[\citet{gibson1997,soubeyrandetal2008}] and
in models for the 2001 UK foot-and-mouth disease epidemic
[\citet{chissterferguson2007}]. %
Recently, \citet{geilhufeetal2012} found that using (fixed)
power-law weights
between regions performed better than real traffic data in predicting influenza
counts in Northern Norway. 
In both models for spatio-temporal surveillance data presented in the
following sections, the power law will be estimated jointly with all
other unknown parameters. Since the choice of a power law is a strong
(yet well motivated) assumption, a comparison with alternative
qualitative formulations is provided.

This paper is organised as follows: in Sections~\ref{sectwinstim} and
\ref{sechhh4},
respectively, the two model frameworks are reviewed and extended with
power-law formulations for the spatial interaction of units. In
Section~\ref{secresults} surveillance data on invasive meningococcal
disease (IMD) and influenza are reanalysed using power laws and alternative
qualitative approaches to be evaluated against previously
used models for these data. We close with some discussion in
Section~\ref{secdiscussion} and a software overview in the \hyperref[appendixsoftware]{Appendix}.
The paper is accompanied by animations (\hyperref[supplemA]{Supplement~A}) and further
supplementary material [Supplement~B: \citet{supplementB}].

\section{Individual-level model} \label{sectwinstim}

\subsection{Introduction}

The spatio-temporal point process model proposed by
\citet{meyeretal2011} is designed for
time--space-mark data $\{(t_i,\mathbf{s}_i,\mathbf{m}_i)\dvtx  i = 1,\ldots,n\}$
of individual case reports
to describe the occurrence of infections (`events') and their
potential to trigger secondary cases.
Formally, the model characterises a point process in
a region $\mathbf{W}$ observed during a period $(0,T]$ through the conditional
intensity function
%
\begin{equation}
\label{eqntwinstim} \lambda(t,\mathbf{s}) = \nu_{[t][\mathbf{s}]} \rho _{[t][\mathbf{s}]} + \sum
_{j\dvtx  t_j < t} \eta_j \cdot g(t-t_j)
\cdot f\bigl(\llVert \mathbf{s}-\mathbf{s}_j\rrVert \bigr).
\end{equation}
%
Related models are the purely temporal, ``self-exciting'' process
proposed by \citet{hawkes1971}, the spatio-temporal epidemic-type
aftershock-sequences (ETAS) model from earthquake research [\citet
{ogata1998}],
the point process models discussed by
\citet{diggle2007}, and an additive-multiplicative point process
model for
discrete-space surveillance data proposed by \citet{hoehle2009}.

The first endemic component in model (\ref{eqntwinstim}) consists of a
log-linear predictor
$\log(\nu_{[t][\mathbf{s}]}) = \beta_0 + \bolds{\beta}^\top\mathbf
{z}_{[t][\mathbf{s}]}$
proportional to an offset $\rho_{[t][\mathbf{s}]}$, typically the population
density. Both the offset and the exogenous covariates are given piecewise
constant on a spatio-temporal
grid (e.g., week${}\times{}$district), hence the notation $[t][\mathbf{s}]$
for the
period which contains $t$ in the region covering $\mathbf{s}$.
In the IMD application in Section~\ref{secresultstwinstim},
$\mathbf{z}_{[t][\mathbf{s}]} = ([t], \sin(\omega\cdot[t]), \cos(\omega
\cdot[t]))^\top$
incorporates a time trend with one sinusoidal wave of frequency
$\omega= 2\pi/365$.

A purely endemic intensity model without the
observation-driven epidemic component is equivalent to a
Poisson regression model for the aggregated number of cases on the chosen
spatio-temporal grid.
However, with an epidemic \mbox{component} the intensity process depends on previously
infected individuals and becomes ``self-exciting.''
Specifically, the epidemic force of infection at $(t,\mathbf{s})$
is the superposition of the infection pressures caused by each
previously infected individual $j$.
The individual infection pressure is weighted by the log-linear predictor
$\log(\eta_j) = \gamma_0 + \bolds{\gamma}^\top\mathbf{m}_j$, which
models effects of individual/infection-specific characteristics $\mathbf{m}_j$
such as the age of the infective.
Regional-level covariates could also be included in $\mathbf{m}_j$, for
example, to model
ecological effects on infectivity.
Note, however, since the epidemic is modelled through a point process, the
susceptible ``population'' consists of the continuous observation region
$\mathbf{W} \subset\mathbb{R}^2$ and is thus infinite.
Consequently, the model cannot include information on
susceptibles, nor an autoregressive term as in time-series models.

Decreasing infection pressure of individual $j$ over space and time
is described by $f(x)$ and $g(t)$, parametric functions of the spatial distance
$x$ and of the elapsed time $t$ since individual $j$ became infectious,
respectively.
The spatial interaction could also be described more generally by a
nonisotropic function $f_2(\mathbf{s})$ of the vector $\mathbf{s}$ to the host,
for example, to incorporate the dominant wind direction in vector-borne
diseases.
However, in our application, $f$ essentially reflects people's
movements and we assume that $f_2(\mathbf{s}) = f(\llVert \mathbf
{s}\rrVert )$ only
depends on
the distance to the host.
Note that we project geographic coordinates into a planar coordinate
reference system to apply Euclidean geometry.
\citet{meyeretal2011} used an isotropic Gaussian kernel
%
\begin{equation}
\label{eqnsiafgaussian} f(x) = \exp \biggl( -\frac{x^2}{2 \sigma
^2} \biggr)
\end{equation}
with scale parameter $\sigma$.
In what follows, we propose an alternative spatial interaction function,
which allows for occasional long-range transmission of infections:
a power law.

\subsection{Power-law extension} \label{secpowerlawtwinstim}

\begin{figure}
\begin{tabular}{@{}ccc@{}}

\includegraphics{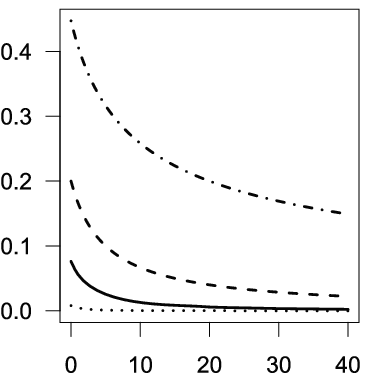}
 & \includegraphics{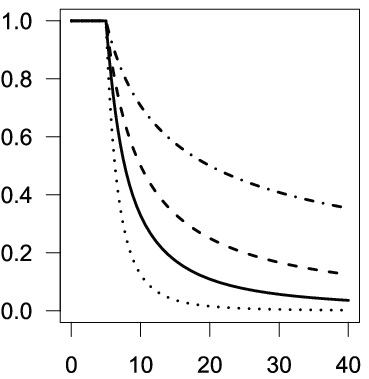} & \includegraphics{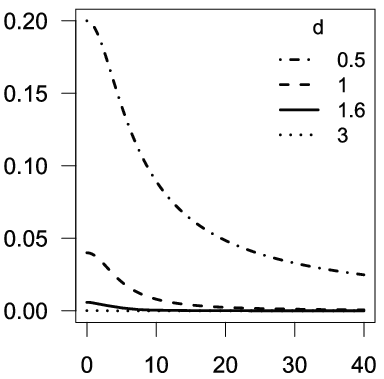}\\
\footnotesize{(a) Power-law kernel} & \footnotesize{(b) Lagged power law} & \footnotesize{(c) Student kernel}
\end{tabular}
\caption{Power-law kernels as a function of the distance $x$
for various choices of the decay parameter $d>0$ and $\sigma=5$.}\label{figPLkernels}
\end{figure}

The basic power law $f(x) = x^{-d}$, $d > 0$, is not a suitable choice for
the distance decay of infectivity since it has a pole at $x=0$.
For $x \ge\sigma> 0$, $x^{-d}$ is the kernel of a Pareto density,
but a shifted version to the domain $\mathbb{R}_0^+$,
known as Pareto type II and sometimes named after \citet{lomax1954},
has density kernel
%
\begin{equation}
\label{eqnsiafpowerlaw} f(x) = (x+\sigma)^{-d} \propto \biggl(1+\frac{x}{\sigma}
\biggr)^{-d}
\end{equation}
%
[see Figure~\ref{figPLkernels}(a)].
Note that there is no need for the spatial interaction function to be normalised
to a density. It is actually more closely related to correlation
functions known from stationary random field models for geostatistical data
[\citet{ChilesDelfiner2012}]. For instance,
the rescaled version $(1+x/\sigma)^{-d}$ is a member of
the Cauchy class introduced by \citet{gneitingschlather2004},
which provides asymptotic power-law correlation as $x \to\infty$.

%

For short-range travel within 10 km,
\citet{brockmannetal2006} found a uniform distribution instead of
power-law behaviour, which suggests an alternative formulation with a ``lagged''
power law:
%
\begin{equation}
\label{eqnsiafpowerlawL} f(x) = \cases{1, &\quad for $x \le\sigma$,
\vspace*{3pt}\cr
\displaystyle
\biggl( \frac{x}{\sigma} \biggr)^{-d}, &\quad otherwise.}
\end{equation}
Spatial interaction is now constant up to the change point $\sigma>0$,
followed by
a power-law decay for larger distances [see Figure~\ref{figPLkernels}(b)].
A similar kernel was used by
\citeauthor{deardonetal2010} [(\citeyear{deardonetal2010}), therein
called ``geometric''] 
for the 2001 UK foot-and-mouth disease epidemic, additionally limiting
spatial interaction to a prespecified upper-bound distance.

A change-point-free kernel also unifying intended short-range and long-range
characteristics is the Student kernel
%
\begin{equation}
\label{eqnsiafstudent} f(x) = \bigl(x^2+\sigma^2
\bigr)^{-d} \propto \biggl(1+ \biggl(\frac{x}{\sigma}
\biggr)^2 \biggr)^{-d}
\end{equation}
with scale parameter $\sigma$ and shape (decay) parameter $d$ [see
Figure~\ref{figPLkernels}(c)].
This kernel implements a power law of the squared distance and is known
as the
`Cauchy model' in geostatistics [\citet{ChilesDelfiner2012}].
For $d > 0.5$, it describes a Student distribution with $2d-1$ degrees of
freedom.

To investigate the appropriateness of the assumed power-law decay, we also
estimate an unconstrained step function 
%
\begin{equation}
\label{eqnsiafstep} f(x) = \sum_{k=0}^K
\alpha_k \mathbh{1}(x \in I_k),
\end{equation}
which corresponds to treating the distance $x$---categorised into consecutive
intervals $I_k$---as a qualitative variable.

\subsection{Inference}

Model parameters are estimated via maximization of the full (log-)likelihood,
applying a quasi-Newton algorithm with analytical gradient and Hessian
[see \citet{supplementB}, Section~1.1].
We estimate kernel parameters on the log-scale to avoid constrained
optimization.
For the step function, $\alpha_0=1$ is fixed to ensure identifiability.

The point process likelihood incorporates the integral of $f_2(\mathbf{s})$
over shifted versions of the observation region $\mathbf{W}$, which is
represented by
polygons.
Similar integrals arise for the partial derivatives of $f_2(\bm{s})$ in the score function and approximate
Fisher information.
Except for the step function kernel (\ref{eqnsiafstep}),
this requires a~method of numerical integration such as the two-dimensional
midpoint rule with an adaptive bandwidth, which was found to be best suited
for the Gaussian kernel [\citet{meyeretal2011}].
For the other kernels we use a more sophisticated approach inspired by
product Gauss cubature over polygons [\citet{sommarivavianello2007}].
This cubature rule is based on Green's theorem, which relates the double
integral over the polygon to a line integral along the polygon boundary.
Its efficiency can be greatly improved in our specific case by taking
analytical advantage of the isotropy of $f_2$, after which numerical
integration remains in only one dimension [see \citet{supplementB}, Section~2.4].
Regardless of any sophisticated cubature rule, the required integration
of $f_2$ over $n$ polygons in the log-likelihood is the part that makes model
fitting cumbersome: it introduces numerical errors which have to be
controlled such that they do not corrupt numerical likelihood
maximization, and
it increases computational cost by several orders of magnitude.
For instance, in our IMD application in Section~\ref{secresultstwinstim}
a single likelihood evaluation would only take 0.02 seconds if we used
a constant
spatial interaction function $f(x) \equiv1$, where the integral does
not depend on
parameters being optimised and simply equals the area of the polygonal domain.
For the Gaussian kernel, a single evaluation takes about 5 seconds, the
step function takes 7 seconds, and the power law and Student kernel
take about
20 seconds.
The above and all following runtime statements refer to total CPU time at
2.80GHz (real elapsed time is shorter since some computations run in
parallel on
multiple CPUs).


\section{Count data model} \label{sechhh4}

\subsection{Introduction}

The multivariate time-series model established by \citet{heldpaul2012}
[see also \citet{paulheld2011,pauletal2008,heldetal2005}]
is designed for spatially and temporally aggregated surveillance data,
that is, disease counts $Y_{it}$ in regions $i=1,\ldots,I$ and periods
$t=1,\ldots,T$.
Formally, the counts $Y_{it}$ are assumed to follow a negative binomial
distribution
\[
Y_{it} \vert\mathbf{Y}_{\cdot,t-1} \sim\NegBin(\mu_{it},
\psi), \qquad i = 1,\ldots,I, t=1,\ldots,T
\]
with additively decomposed mean
%
\begin{equation}
\label{eqnmeanHHH} \mu_{it} = \nu_{it} e_{it} +
\lambda_{it} Y_{i,t-1} + \phi_{it} \sum
_{j
\ne i} w_{ji} Y_{j,t-1},
\end{equation}
and overdispersion parameter $\psi$ such that the conditional variance of
$Y_{it}$ is $\mu_{it} (1+\psi\mu_{it})$. The Poisson distribution
results as
a special case if $\psi= 0$.
In (\ref{eqnmeanHHH}), the first term represents the endemic component
similar to the point process model (\ref{eqntwinstim}). The endemic
mean is
proportional to an offset of known expected counts $e_{it}$ typically
reflecting the population at risk.
The other two components are observation-driven epidemic components:
an autoregression on the number of cases at the previous time point,
and a ``spatio-temporal'' component capturing transmission from
other units. Note that without these epidemic components, the model
would reduce
to a negative binomial regression model for independent observations.

Each of $\nu_{it}$, $\lambda_{it}$, and $\phi_{it}$ is a
log-linear predictor of the form
\[
\log(\cdot_{it}) = \alpha^{(\cdot)} + b_i^{(\cdot)}
+ {\bolds{\beta}^{(\cdot)}}^\top\mathbf{z}^{(\cdot)}_{it}
\]
(where ``$\cdot$'' is one of $\nu$, $\lambda$, $\phi$),
containing fixed and region-specific intercepts as well as effects
of exogenous covariates $\mathbf{z}^{(\cdot)}_{it}$ including time effects.
For example, in the influenza application in Section~\ref{secresultshhh4},
\[
\mathbf{z}^{(\nu)}_{it} = \bigl(t, \sin(1 \cdot\omega t), \cos(1
\cdot\omega t), \ldots, \sin(S \cdot\omega t), \cos(S \cdot\omega t)
\bigr)^\top
\]
describes an endemic time trend with a superposition of $S$ harmonic
waves of
fundamental frequency $\omega=2\pi/52$ [\citet{heldpaul2012}].
The random effects $\mathbf{b}_i:= (b_i^{(\lambda)}, b_i^{(\phi)},
b_i^{(\nu)})^\top$ account for heterogeneity between regions, and are
assumed to follow independently a trivariate normal distribution with
mean zero and covariance matrix $\bolds{\Sigma}$. Accounting for
correlation of random effects across regions is possible by
adopting a conditional autoregressive (CAR) model
[\citet{paulheld2011}].

The weights $w_{ji}$ of the spatio-temporal
component in (\ref{eqnmeanHHH}) describe the strength of transmission
from region $j$ to region $i$, collected into an $I \times I$ weight matrix
$(w_{ji})$. 
In contrast to the individual-level model, all of the $Y_{j,t-1}$
cases of the neighbour $j$ by aggregation contribute with the same weight
$w_{ji}$ to infections in region~$i$.
In previous work, these weights were assumed to be known and restricted to
first-order neighbours:
%
\begin{equation}
\label{eqnwji1norm} w_{ji} = \cases{ 1/n_j, &\quad for $i \sim
j$,
\vspace*{3pt}\cr
0, &\quad otherwise,}
\end{equation}
where the symbol ``$\sim$'' denotes ``is adjacent to'' and $n_j$ is the number
of direct (first-order) neighbours of region $j$. This is a normalised version
of the ``raw'' adjacency indicator matrix
$\mathbf{A} = (\mathbh{1}(i \sim j))_{j,i = 1,\dots,I}$, which is binary
and symmetric.
The idea behind normalisation is that each region $j$ distributes its cases
uniformly to its $n_j$ neighbours [\citet{pauletal2008}].
Accordingly, the weight matrix is normalised to proportions such
that all rows sum to 1.
A simple alternative weight matrix considering only first-order
neighbours would
result from the definition $w_{ji} = 1/n_i$ for $i\sim j$ (i.e.,
columns sum to
1), meaning that the number of cases in a region $i$ at time $t$ is
promoted by
the mean of the neighbours at time $t-1$. However, the
first definition seems more natural in the framework of branching processes,
where the point of view is from the infective source. Furthermore, the factor
$1/n_i$ would be confounded with the region-specific effects
$b_i^{(\phi)}$.

In either case, with the above weight matrix, the epidemic can only
spread to first-order neighbours during the period $t \rightarrow t+1$,
except for independently imported cases via the endemic component.
This ignores the ability of humans to travel further.
In what follows, we propose a parametric generalisation of the neighbourhood
weights: a power law.

\subsection{Power-law extension} \label{secpowerlawhhh4}

To implement the power-law principle in the network of geographical
regions, we first need to define a distance measure on which the
power law acts.
There are two natural choices:
Euclidean distance between centroid coordinates
and the order of neighbourhood.
The first one conforms to a continuous power law, whereas the second
one is
discrete.
However, using centroid coordinates interferes with the area and shape
of the
regions. Specifically, a tiny neighbouring region would be attributed
a stronger link than a large neighbour with centroid further apart,
even if the latter shares more boundary than the tiny region.
Using the common boundary length as a measure of ``coupling''
[\citet{keelingrohani2002}] would only cover adjacent regions.
We thus opt for the discrete measure of neighbourhood order.

Formally, a region $j$ is a $k$th-order neighbour of another region $i$,
denoted $o_{ji} = o_{ij} = k$, if it is adjacent to a $(k-1)$th-order neighbour
of $i$ and if it is not itself a neighbour of order $k-1$ of region $i$.
In other words, two regions are $k$th-order neighbours, if the shortest route
between them has $k$ steps across distinct regions.
The network of regions thus features
a symmetric $I \times I$ matrix of neighbourhood orders
with zeroes on the diagonal by convention.


Given this discrete distance measure, we generalise the previously used
first-order weight matrix to higher-order neighbours assuming a power
law with
decay parameter $d > 0$:
%
\begin{equation}
\label{eqnwjiraw} w_{ji} = o_{ji}^{-d}
\end{equation}
for $j \ne i$ and $w_{jj} = 0$. This may also be recognised as the kernel
of the \citet{Zipf1949} probability distribution.
The raw power-law weights (\ref{eqnwjiraw}) can be normalised to
%
\begin{equation}
\label{eqnwjinorm} w_{ji} = \frac{o_{ji}^{-d}}{\sum_{k=1}^I o_{jk}^{-d}}
\end{equation}
such that $\sum_{k=1}^I w_{jk} = 1$ for all rows $j$ of the weight matrix.
The higher the decay parameter $d$, the less important are higher-order
neighbours. The limit $d \rightarrow\infty$ corresponds to the
previously used
first-order dependency, whereas $d=0$ would assign equal weight to all regions.

Similarly to the point process modelling in Section~\ref{secpowerlawtwinstim}, we also
estimate the weights in a qualitative way by treating
the order of neighbourhood as a factor:
%
\begin{equation}
\label{eqnwjiunconstrained} w_{ji} = \sum_{o=1}^{M-1}
\omega_o \cdot\mathbh{1}(o_{ji} = o) +
\omega_M \cdot\mathbh{1}(o_{ji} \ge M).
\end{equation}
Aggregation of higher orders ($o_{ji} \ge M$) is necessary since the
available information becomes increasingly sparse. As before, the
unconstrained weights (\ref{eqnwjiunconstrained}) can be normalised
to $w_{ji} / \sum_{k=1}^I w_{jk}$.

\subsection{Inference}

We set $\omega_1 = 1$ for identifiability and estimate the decay
parameter $d$
and the unconstrained weights $\omega_2,\ldots,\omega_M$ on the
log-scale to
enforce positivity.
Supplied with the enhanced score function and Fisher information matrix,
estimation of parametric weights is still possible within the penalised
likelihood framework established by \citet{paulheld2011}
[see also \citet{supplementB}, Section~1.2].
The authors argue, however, that classical model choice criteria such as
Akaike's Information Criterion (AIC) cannot be used straightforwardly for
models with random effects.
Therefore, performance of the power-law models and the previous first-order
formulations is compared by one-step-ahead forecasts assessed
with strictly proper scoring rules:
the logarithmic score (logS) and the ranked probability
score (RPS) advocated by \citet{czado-etal-2009} for count data:
\begin{eqnarray*}
\operatorname{logS}(P,y) &=& - \log P(Y=y),
\\
\operatorname{RPS}(P,y) &=& \sum_{k=0}^\infty
\bigl[P(Y\le k) - \mathbh{1}(y\le k) \bigr]^2.
\end{eqnarray*}
These scores evaluate the discrepancy between the predictive
distribution $P$ from a fitted model and the later observed value
$y$. Thus, lower scores correspond to better predictions. Note that
the infinite sum in the RPS can be approximated by truncation at some
large $k$ in a way such that a prespecified absolute approximation
error is maintained [\citet{weiheld2013}]. Such scoring rules have
already been used for previous analyses of the influenza surveillance
data [\citet{heldpaul2012}]. Along these lines, one-step-ahead
predictions and associated scores are computed and statistical
significance of the difference in mean scores is assessed using a
Monte-Carlo permutation test for paired data.

\section{Applications} \label{secresults}

We now apply the power-law formulations of both model frameworks to previously
analysed surveillance data and investigate potential improvements with respect
to predictive performance. We investigate the appropriateness of the power-law
shape by alternative qualitative estimates of spatial \mbox{interaction}.
In Section~\ref{secresultstwinstim}
635 individual case reports of IMD caused by the two
most common bacterial finetypes of meningococci in Germany from 2002 to 2008
are analysed with the point process model (\ref{eqntwinstim}).
In Section~\ref{secresultshhh4} the multivariate time-series model
(\ref{eqnmeanHHH}) is applied to weekly numbers of reported cases of influenza
in the 140 administrative districts of the federal states Bavaria and
Baden-W\"urttemberg in Southern Germany from 2001 to 2008.
In Section~\ref{secresultsflusims} we evaluate a simulation-based
long-term forecast
of the 2008 influenza wave.
Space--time animations of both surveillance data sets are provided in
\hyperref[supplemA]{Supplement~A}.

\begin{figure}[t]
\begin{tabular}{@{}c@{\quad}c@{}}

\includegraphics{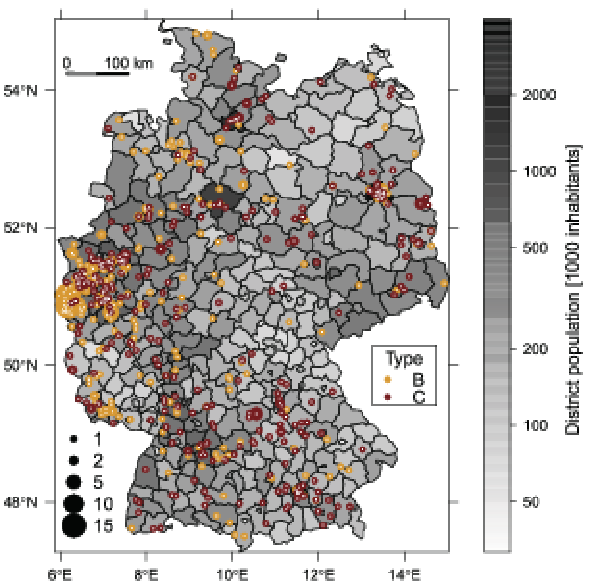}
 & \includegraphics{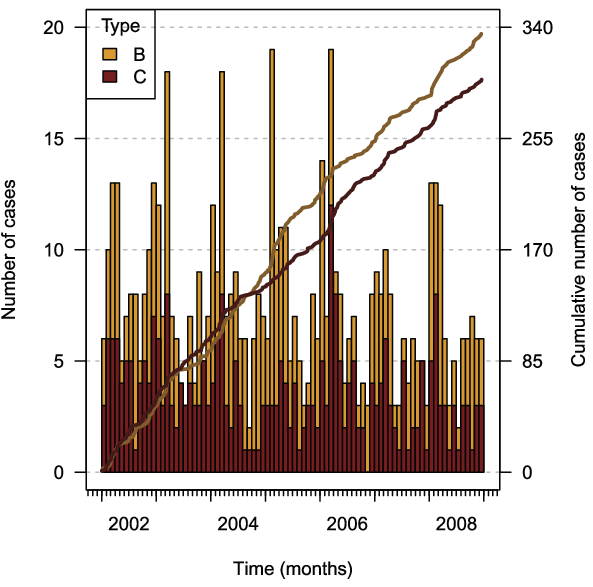}\\
\footnotesize{(a) Spatial point pattern with dot size}  & \footnotesize{(b) Monthly aggregated time series and}\\[-1.5pt]
\footnotesize{proportional to the number of cases at}   & \footnotesize{evolution of the cumulative number of cases}\\[-1.5pt]
\footnotesize{the respective location (postcode level)} & \footnotesize{(by date of specimen sampling)}
\end{tabular}
\caption{Distribution of the 635 IMD cases in Germany,
2002--2008, caused by the two most common meningococcal finetypes
\mbox{B:P1.7-2,4:F1-5}
(335 cases) and
\mbox{C:P1.5,2:F3-3}
(300 cases),
as reported to and typed by the German Reference Centre for Meningococci.}
\label{figIMD}
\end{figure}

\subsection{Cases of invasive meningococcal disease in Germany, 2002--2008 (see Figure~\texorpdfstring{\protect\ref{figIMD}}{2})}\label{secresultstwinstim}

In the original analysis of the IMD data [\citet{meyeretal2011}],
comprehensive AIC-based model selection yielded a linear time trend, a
sinusoidal time-of-year effect ($S=1$), and no effect of the (lagged)
number of
local influenza cases in the endemic component.
The epidemic component included an effect of the meningococcal finetype
(\mbox{C:P1.5,2:F3-3} being less infectious than \mbox{B:P1.7-2,4:F1-5},
abbreviated by C and B in the following), a small age effect (\mbox{3--18} year old
patients tending to be more infectious), and supported an isotropic Gaussian
spatial interaction function $f$ compared to a homogeneous spatial spread
[$f(x)\equiv1$].
The analysis assumed constant infectivity over
time until 30 days after infection when infectivity vanishes to zero,
that is, $g(t)=\mathbh{1}_{(0,30]}(t)$.
In this paper, we replace the Gaussian kernel in the selected model by the
proposed power-law distance decay (\ref{eqnsiafpowerlaw}) to
investigate if it
better captures the dynamics of IMD spread.

%

Note that the distinction between two finetypes in this application actually
corresponds to a marked version of the point process model.
It is described by an intensity function $\lambda(t,\mathbf{s},k)$, where the
sum in (\ref{eqntwinstim}) is restricted to previously infected individuals
with bacterial finetype $k$, since we assume that infections of
different finetypes are not associated via transmission [\citet
{meyeretal2011}].
For convenience, we kept notation simple and comparable to the multivariate
time-series model of Section~\ref{sechhh4}.

Prior to fitting point process models to the IMD data,
the interval-censored nature of the data caused by a restricted resolution
in space and time has to be taken into account:
we only observed dates and residence postcodes of the cases
(implicitly assuming that infections effectively happened within the
residential neighbourhood).
This makes the data interval-censored, yielding tied observations.
However, ties are not compatible with our (continuous-time, continuous-space)
point process model since observing two events at the exact same time
point or
location has zero probability.
In the original analyses with a Gaussian kernel $f$, events were untied
in time
by subtracting a $U(0,1)$-distributed random number from all observed time
points [\citet{meyeretal2011}], that is, random sampling within
each day,
which is also the preferred method used by \citet{diggleetal2009}.
To identify the two-parameter power law $(x+\sigma)^{-d}$,
it was additionally necessary to break ties in space, since otherwise
$\log\sigma$
diverged to $-\infty$, yielding a pole at $x=0$.
A possible solution is to shift all locations randomly in space within their
round-off intervals similar to the tie-breaking in time. Lacking a
shapefile of the postcode regions, we shifted locations
by a vector uniformly drawn from the disc with radius $\varepsilon/2$, where
$\varepsilon$ is the minimum observed spatial separation of distinct
points, here $\varepsilon=1.17$ km.
Accordingly, a sensitivity analysis was conducted by applying the random
tie-breaking in time and space 30 times and fitting the models to all
replicates.

%
\begin{figure}[b]
\begin{tabular}{@{}cc@{}}

\includegraphics{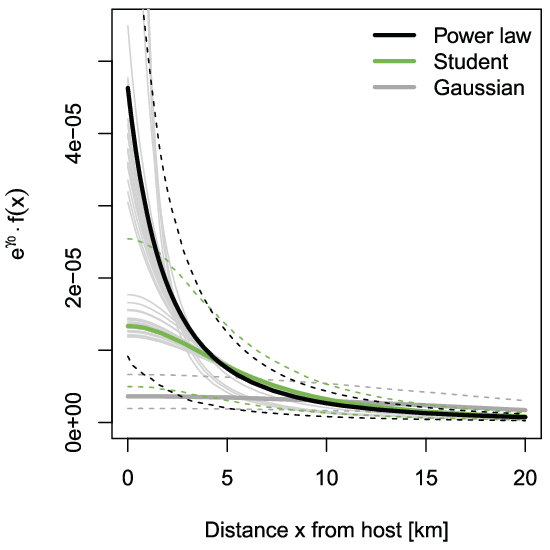}
 & \includegraphics{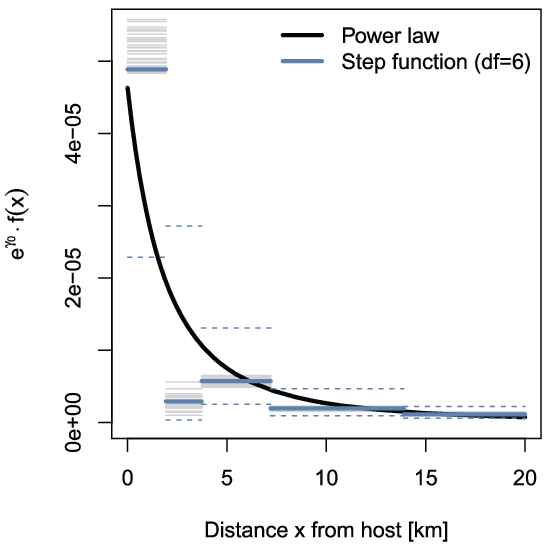}\\
\footnotesize{(a) Power laws vs. the Gaussian kernel}  & \footnotesize{(b) Power law vs. a step function}
\end{tabular}
\caption{Estimated spatial interaction functions---appropriately
scaled by the
epidemic intercept $\exp(\gamma_0)$.
The dashed lines represent 95\% confidence intervals obtained as the pointwise
2.5\% and 97.5\% quantiles of the functions evaluated for 999 samples
from the
asymptotic multivariate normal distribution of the affected parameters.
The light grey lines are estimates obtained from a sensitivity analysis
with repeated random tie-breaking.}\label{figsiafs}
\end{figure}

%
\begin{table}[t]
\tabcolsep=0pt
\caption{Parameter estimates and 95\% Wald confidence intervals for the
Gaussian and the power-law model.
Results for the Gaussian kernel are slightly different from those
reported by \citet{meyeretal2011} due to improved numerical integration.
Note that we use the symbol $\sigma$ for the scale parameter and $d$
for the decay parameter in all spatial interaction functions, but these
parameters as well as $\gamma_0$ are not directly comparable (instead see
Figure~\protect\ref{figsiafs})}\label{tabcoefs}
\begin{tabular*}{\tablewidth}{@{\extracolsep{\fill}}@{}ld{3.2}d{3.11}d{3.2}d{3.11}@{}}
\hline
& \multicolumn{2}{c}{\textbf{Gaussian kernel (\ref{eqnsiafgaussian})}} & \multicolumn{2}{c@{}}{\textbf{Power-law kernel (\ref{eqnsiafpowerlaw})}} \\[-6pt]
& \multicolumn{2}{c}{\hrulefill} & \multicolumn{2}{c@{}}{\hrulefill}\\
& \multicolumn{1}{c}{\textbf{Estimate}} & \multicolumn{1}{c}{\textbf{95\% CI}}
& \multicolumn{1}{c}{\textbf{Estimate}} & \multicolumn{1}{c@{}}{\textbf{95\% CI}}\\
\hline
$\beta_0$ & -20.53 &  {-}20.62  \mbox{ to }  {-}20.44 & -20.58 &  {-}20.68  \mbox{ to }  {-}20.47 \\
$\beta_{\mathrm{trend}}$ & -0.05 &  {-}0.09  \mbox{ to }  {-}0.00 & -0.05 &  {-}0.09  \mbox{ to }  0.00 \\
$\beta_{\sin}$ & 0.26 &  0.14  \mbox{ to }  0.39 & 0.26 &  0.12  \mbox{ to }  0.39 \\
$\beta_{\cos}$ & 0.26 &  0.14  \mbox{ to }  0.39 & 0.27 &  0.14  \mbox{ to } 0.40 \\
$\gamma_0$ & -12.53 &  {-}13.15  \mbox{ to }  {-}11.91 & -6.21 &  {-}9.32  \mbox{ to } {-}3.10 \\
$\gamma_{\mathrm{C}}$ & -0.91 &  {-}1.44  \mbox{ to }  {-}0.39 & -0.80 &  {-}1.31  \mbox{ to }  {-}0.29 \\
$\gamma_{3{-}18}$ & 0.67 &  0.04  \mbox{ to }  1.31 & 0.78 &  0.11  \mbox{ to }  1.45 \\
$\gamma_{\geq19}$ & -0.29 &  {-}1.19  \mbox{ to }  0.61 & -0.18 &  {-}1.11 \mbox{ to }  0.75 \\
$\sigma$ & 16.37 &  13.95  \mbox{ to }  19.21 & 4.60 &  1.80  \mbox{ to }  11.71 \\
$d$ &  &  &  2.47 &  1.80  \mbox{ to }  3.39 \\
\hline
\end{tabular*}\vspace*{-6pt}
\end{table}

Figure~\ref{figsiafs}(a) displays estimated spatial interaction functions---appropriately scaled by $\exp(\hat\gamma_0)$---together with confidence intervals and estimates from the sensitivity analysis
(see Table~\ref{tabcoefs} for values of $\hat\gamma_0$, $\hat
\sigma$, and $\hat d$).
The power law puts much more weight on localised
transmissions with an initially faster distance decay of infectivity.
Furthermore, it features a heavier tail than the Gaussian kernel, which
facilitates the geographical spread of IMD by occasional long-range
transmissions. Maps of the accumulated epidemic intensity
[\citet{supplementB}, Figure~1]
visualise the impact of the power law on the modelled infectivity.
Sensitivity analysis shows that AIC clearly prefers the new power-law kernel
against the Gaussian kernel (mean $\Delta\operatorname{AIC}=-27.6$,
$\sd=1.5$).
The Student kernel represents a compromise between the other two
parametric kernels with short-range properties similar to the Gaussian kernel
but with a heavy tail. However, AIC improvement is not as large as for the
above power law (mean $\Delta\operatorname{AIC}=-15.5$, $\sd=0.9$).

For these three kernels, sensitivity analysis of the random tie-breaking
procedure in space and time generally confirmed the results. The
Gaussian kernel
was least affected by the small-scale perturbation of event times and locations.
Some replicates for the power-law model yielded a slightly steeper shape,
which is due to closely located points after random tie-breaking.
Such an artifact would have
been avoided if we had used constrained sampling in that the randomly shifted
points obey a minimum separation of say 0.1 km.

The estimated lagged version of the power law (\ref{eqnsiafpowerlawL})
is shown in Supplement~B [\citet{supplementB}, Figure~2]. It
has
a uniform short-range dispersal radius
of $\hat{\sigma}=0.40$ (95\% CI: 0.18 to 0.86)
kilometres. However, such a small $\sigma$ is not interpretable since
it is
actually not covered by the spatial resolution of the data.
Accordingly, the 30 estimates of the sensitivity analysis are
more dispersed, as is the goodness of fit compared to the Gaussian kernel
(mean $\Delta\operatorname{AIC}=-21.1$, $\sd=3.8$).

Figure~\ref{figsiafs}(b) shows a comparison of the estimated power law
with a step function~(\ref{eqnsiafstep}) for spatial interaction.
An upper boundary knot had to be specified, which we set at 100 kilometres,
where the step function drops to 0.
We chose six knots to be equidistant on the log-scale within $[0,\log(100)]$,
that is, steps at 1.9, 3.7, 7.2, 13.9, 26.8, and 51.8
kilometres. Estimation took only 72
seconds due to the analytical implementation of the integration of
$f_2$ over polygonal domains, whereas the power-law model took
42 minutes.
The power law is well confirmed by the step function; it is almost completely
enclosed by its 95\% confidence interval.
The step function suggests an even steeper initial decay and has a slightly
better fit in terms of AIC (mean $\Delta\operatorname{AIC}=-6.9$,
$\sd=4.0$ compared to
the power law).
However, it depends on the choice of knots, it is sensitive for artifacts
of the data and forfeits monotonicity, which contradicts
Tobler's \emph{first law of geography} [\citet{tobler1970}].


Parameter estimates and confidence intervals for the Gaussian and the power-law
model are presented in Table~\ref{tabcoefs}
[see \citet{supplementB}, Table~1, for parameter estimates of the
other models].
The parameters of the endemic component characterising time trend and
seasonality were not affected by the change of the shape of spatial interaction,
and also the epidemic coefficients of finetype and age group do not
differ much
between the models retaining their signs and orders of magnitude.
For instance, also with the power-law kernel,
the C-type is approximately half as infectious as the B-type, which is
estimated by the multiplicative type-effect
$\exp(\hat\gamma_{\mathrm{C}}) =0.45$ (95\% CI: 0.27 to 0.75)
on the force of infection (type~B is the reference category here).

An important quantity in epidemic modelling is the expected number $R$
of offspring (secondary infections) each case generates.
This reproduction number can be derived from the fitted models for each
event by
integrating its triggering function $\eta_j g(t-t_j) f(\llVert
\mathbf {s}-\mathbf{s}_j\rrVert )$
over the observation region $\mathbf{W}$ and period $[t_j,T]$ [\citet
{meyeretal2011}].
Type-specific estimates of $R$ are then obtained by averaging over
the individual estimates by finetype.
Table~\ref{tabR0s} shows that the reproduction numbers become
slightly larger,
which is related to the heavier tail of the power law enabling
additional interaction between events at far distances.

%
\begin{table}[b]
\caption{Type-specific reproduction numbers with 95\% confidence
intervals (based on 199 samples from the asymptotic multivariate normal
distribution of the parameter estimates)}\label{tabR0s}
\begin{tabular*}{\tablewidth}{@{\extracolsep{\fill}}@{}lcccc@{}}
\hline
& \multicolumn{2}{c}{\textbf{Gaussian kernel (\ref{eqnsiafgaussian})}} & \multicolumn{2}{c@{}}{\textbf{Power-law kernel (\ref{eqnsiafpowerlaw})}} \\[-6pt]
& \multicolumn{2}{c}{\hrulefill} & \multicolumn{2}{c@{}}{\hrulefill}\\
& \multicolumn{1}{c}{\textbf{Estimate}} & \multicolumn{1}{c}{\textbf{95\% CI}}
& \multicolumn{1}{c}{\textbf{Estimate}} & \multicolumn{1}{c@{}}{\textbf{95\% CI}}\\
\hline
B & 0.22 & 0.17 to 0.31 & 0.26 & 0.10 to 0.35 \\
C & 0.10 & 0.06 to 0.15 & 0.13 & 0.05 to 0.19 \\
\hline
\end{tabular*}
\end{table}

We close this application with two additional ideas for improvement of the
model.
First, it might be worth considering a population effect also in the
\emph{epidemic} component to reflect higher contact rates and thus infectivity
in regions with a denser population.
Using the log-population density of the infective's district,
$\log\rho_{[t_j][\mathbf{s}_j]}$, the corresponding parameter is
estimated to be
$\hat\gamma_{\log(\rho)} =0.21$ (95\%~CI: $-$0.07 to 0.48), that is, individual infectivity scales
with $\rho^{
0.21
}$, where $\rho$ ranges from
39 to 4225 km$^2$.
Although the positive point estimate supports this idea, the wide confidence
interval does not reflect strong evidence for such a population effect
in the
IMD data.

%
\begin{figure}[b]
\begin{tabular}{@{}cc@{}}

\includegraphics{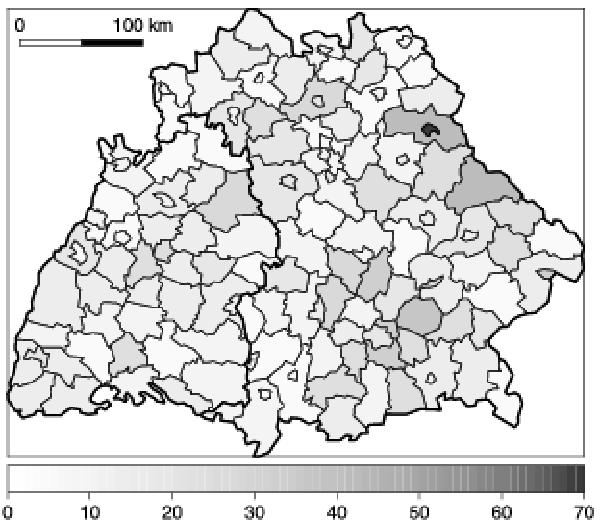}
 & \includegraphics{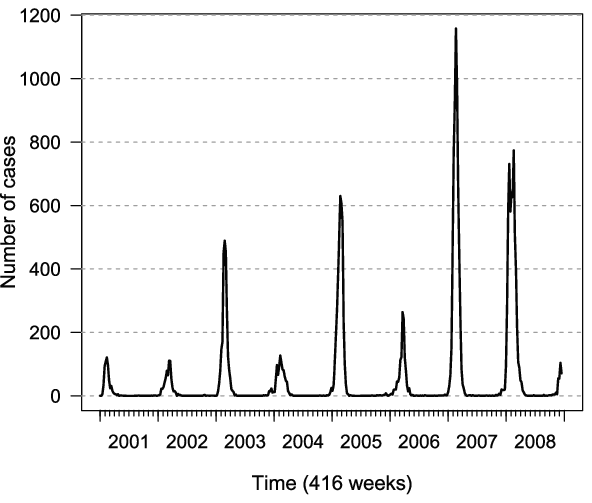}\\
\footnotesize{(a) Mean yearly incidence per 100,000}  & \footnotesize{(b) Weekly number of cases}\\[-1.5pt]
\footnotesize{inhabitants}
\end{tabular}
\caption{Spatial and temporal distribution of reported influenza cases
in the 140 districts of Bavaria and Baden-W\"urttemberg during the years 2001
to 2008.}\label{figflu}
\end{figure}

However, it is helpful to allow for spatial heterogeneity in the \emph{endemic}
component.
For instance, an indicator for districts at the border or the distance
of the district's centroids from the border could serve as proxies for simple
edge effects. The idea is that as we get closer to the edge of the observation
window (Germany) more infections will originate from external sources not
directly linked to the observed history of the epidemic within Germany.
We thus model a spatially varying risk of importing cases through the endemic
component. For the Greater Aachen Region in the central-west part of
Germany, where a spatial disease cluster is apparent in Figure~\ref{figIMD}(a),
such a cross-border effect with the Netherlands was indeed identified by
\citet{eliasetal2010} for the serogroup B finetype during our observation
period using molecular sequence typing of bacterial strains in infected patients
from both countries.
Inclusion of an edge indicator in the endemic covariates $\mathbf
{z}_{[t][\mathbf{s}]}$
improves AIC by
5 
with an estimated rate ratio of
1.37 (95\% CI: 1.10 to 1.70)
for districts at the border versus inner districts.
If we instead use the distance to the border, AIC improves by
20 
with an estimated risk reduction of
5.0\% (95\% CI: 3.0\% to 7.0\%)
per 10 km increase in distance to the border.

\subsection{Influenza surveillance data from Southern Germany, 2001--2008 (see Figure~\texorpdfstring{\protect\ref{figflu}}{4})}\label{secresultshhh4}

The best model (with respect to logS and RPS) for the influenza surveillance
data found by \citet{heldpaul2012} using normalised first-order weights
included $S=1$ sinusoidal wave in each of the
autoregressive ($\lambda_{it}$) and spatio-temporal ($\phi_{it}$) components
and $S=3$ harmonic waves with a linear trend in the endemic component
$\nu_{it}$ with the population fraction $e_{i}$ in region $i$ as offset.
We now fit an extended model by estimating (raw or normalised) power-law
neighbourhood weights (\ref{eqnwjiraw}) or (\ref{eqnwjinorm}) as described
in Section~\ref{secpowerlawhhh4}, which replace the previously used
fixed adjacency
indicator.

%
\begin{figure}[b]
\begin{tabular}{@{}cc@{}}

\includegraphics{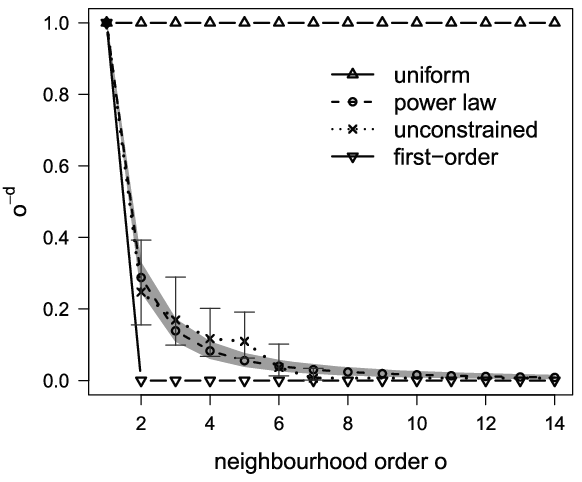}
 & \includegraphics{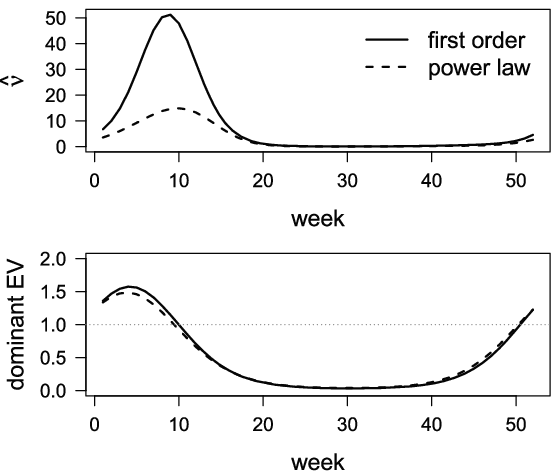}\\
\footnotesize{(a) Power-law (\protect\ref{eqnwjinorm}) and unconstrained}  & \footnotesize{(b) Seasonal variation of the endemic (top)}\\[-1.5pt]
\footnotesize{weights (\protect\ref{eqnwjiunconstrained}) with 95\% confidence intervals}  & \footnotesize{and epidemic (bottom) components}
\end{tabular}
\caption{Estimated power-law and unconstrained weights \textup{(a)}, and seasonal
variation \textup{(b)} using normalised weights.}\label{figfluPL}\label{figfluSeason}
\end{figure}

Figure~\ref{figfluPL}(a) shows the estimated normalised power law with
$\hat{d}=1.80$ (95\% CI: 1.61 to 2.01).
This decay is remarkably close to the power-law exponent 1.59 estimated by
\citet{brockmannetal2006} for short-time travel in the USA with
respect to
distance (in kilometres), even though neighbourhood order is a discretised
measure with no one-to-one correspondence to Euclidean distances,
and travel behaviour in the USA is potentially
different from that in Southern Germany.
The plot also shows the estimated unconstrained weights for comparison
with the
power law. The sixth order of neighbourhood was the highest for which
we could
estimate an individual weight; higher orders had to be aggregated corresponding
to $M=7$ in (\ref{eqnwjiunconstrained}).
The unconstrained weights decrease monotonically and resemble nicely the
estimated power law, which is enclosed by the 95\% confidence
intervals (except for order 5, which has a slightly higher weight).
The results with raw weights are very similar and shown in
\citeauthor{supplementB} [(\citeyear{supplementB}), Figure~3].

Figure~\ref{figfluSeason}(b) shows the estimated seasonal variation
in the endemic
component and the course of the
dominant eigenvalue [\citet{heldpaul2012}] for the normalised
weight models.
The dominant eigenvalue is a combination of the two epidemic components:
if it is smaller than 1, it can be interpreted as the epidemic
proportion of
total disease incidence, otherwise it indicates an outbreak period.
Whereas the course of this combined measure is more or less unchanged upon
accounting for higher-order neighbours with a power law,
the weight of the endemic component decreases remarkably. This goes
hand in hand
with an increased importance of the spatio-temporal component since in the
power-law formulation much more information can be borrowed from the
number of
cases in other regions. Jumps of the epidemic to nonadjacent regions
within one
week are no longer dedicated to the endemic component only.

Concerning the remaining coefficients, there is less overdispersion in the
power-law models (see $\psi$ in Table~\ref{tabfluEstimates}),
which indicates reduced residual heterogeneity.
For the variance and correlation
estimates of the random effects, there is no substantial difference between
first-order and power-law models and even less between raw and normalised
formulations.

%
\begin{table}[b]
\tabcolsep=0pt
\caption{Estimated model parameters (with standard errors) excluding
intercepts and trend/seasonal coefficients. The parameter
$\beta^{(\phi)}_{\mttt{log(pop)}}$ in the first row belongs to a
further extended power-law (PL) model, which accounts for population in the
spatio-temporal component (last column).
The $\sigma^2_{\cdot}$~and~$\rho_{\cdot\cdot}$ parameters are the variances
and correlations of the random effects (from $\bolds{\Sigma}$).
The last row shows the final values of the penalised and marginal
log-likelihoods}\label{tabfluEstimates}
\begin{tabular*}{\tablewidth}{@{\extracolsep{\fill}}@{}lccccc@{}}
\hline
& \multicolumn{2}{c}{\textbf{Raw weights}} & \multicolumn{3}{c@{}}{\textbf{Normalised weights}} \\[-6pt]
& \multicolumn{2}{c}{\hrulefill} & \multicolumn{3}{c@{}}{\hrulefill}\\
& \multicolumn{1}{c}{\textbf{First order}} & \multicolumn{1}{c}{\textbf{Power law}} & \multicolumn{1}{c}{\textbf{First order}}
& \multicolumn{1}{c}{\textbf{Power law}} & \multicolumn{1}{c}{\textbf{PL$\bolds{{}+{}}$pop.}} \\
\hline
${\beta}^{(\phi)}_{\mttt{log(pop)}}$ & \multicolumn{1}{c}{--} & \multicolumn{1}{c}{--} & \multicolumn{1}{c}{--} & \multicolumn{1}{c}{--} & 0.76 (0.13) \\
${d}$ & \multicolumn{1}{c}{--} & 1.72 (0.10) & \multicolumn{1}{c}{--} & 1.80 (0.10) & 1.65 (0.10) \\
${\psi}$ & 0.93 (0.03) & 0.86 (0.03) & 0.92 (0.03) & 0.86 (0.03) & 0.86 (0.03) \\
${\sigma}^2_\lambda$ & 0.14 & 0.17 & 0.13 & 0.17 & 0.16 \\[3pt]
${\sigma}^2_\phi$ & 0.94 & 0.92 & 0.98 & 0.89 & 0.71 \\[3pt]
${\sigma}^2_\nu$ & 0.50 & 0.67 & 0.51 & 0.67 & 0.66 \\
${\rho}_{\lambda\phi}$ & 0.02 & 0.20 & 0.03 & 0.21 & 0.13 \\
${\rho}_{\lambda\nu}$ & 0.11 & 0.31 & 0.12 & 0.31 & 0.27 \\
${\rho}_{\phi\nu}$ & 0.56 & 0.29 & 0.55 & 0.30 & 0.39 \\
$l_{\mathrm{pen}} (l_{\mathrm{mar}})$ & $-$18,400 ($-$433) & $-$18,129 ($-$456)
& $-$18,387 ($-$436) & $-$18,124 ($-$453) & $-$18,124 ($-$439) \\
\hline
\end{tabular*}
\end{table}

To assess if the power-law formulation improves the previous
first-order model, their predictive performance is compared based on
one-week-ahead predictions for all 140 regions and the
104 weeks of the last two years.
Computing these predictions for one model takes about 3 hours,
since it needs to be refitted for every time point.
Table~\ref{tabfluScores} shows the resulting mean scores with
associated $p$-values.
Both logS and RPS improve when
accounting for higher-order neighbours with a power law, while the
difference is only significant for the logarithmic score.
Furthermore, the normalised formulation performs slightly
better than the raw weights. For instance, the mean difference in the
logarithmic scores of the respective power-law models has an associated
$p$-value of 0.0009.
In the following we therefore only consider the normalised versions.
For additional comparison, the simple uniform weight model ($w_{ji}
\equiv1$),
which takes into account higher-order neighbours but with equal weight, has
mean $\operatorname{logS}=0.5484$ and
mean $\operatorname{RPS}=0.4215$,
and thus performs worse than a power-law decay and, according to the
RPS, even
worse than first-order weights.

%
\begin{table}
\tabcolsep=0pt
\caption{Mean scores of $104 \times140$
one-week-ahead predictions over the last two years, accompanied with
$p$-values for comparing power-law and first-order weights obtained via
permutation tests with 19,999 random permutations. Note that the
values obtained for normalised first-order weights are slightly different
from the ones published by \citet{heldpaul2012} due to a
correction of a recording error in the last week of the influenza data}\label{tabfluScores}
\begin{tabular*}{\tablewidth}{@{\extracolsep{\fill}}@{}lcccc@{}}
\hline
& \multicolumn{2}{c}{\textbf{Raw weights}} & \multicolumn{2}{c@{}}{\textbf{Normalised weights}}\\[-6pt]
& \multicolumn{2}{c}{\hrulefill} & \multicolumn{2}{c@{}}{\hrulefill}\\
& \textbf{logS} & \textbf{RPS} & \textbf{logS} & \textbf{RPS} \\
\hline
First order & 0.5522\phantom{0} & 0.4205 & 0.5511 & 0.4194 \\
Power law & 0.5453\phantom{0} & 0.4174 & 0.5448 & 0.4168 \\
$p$-value & 0.00005 & 0.11\phantom{00} & 0.0001 & 0.19\phantom{00} \\
\hline
\end{tabular*}
\end{table}

Similarly to the IMD analysis, further improvement of the model's
description of human mobility can be achieved by accounting for the
district-specific population also in the spatio-temporal component.
The idea is that there tends to be more traffic to regional
conurbations, that is,
districts with a larger population, which are thus expected to
import a bigger amount of cases from\vspace*{1pt} neighbouring regions [\citet
{bartlett1957}].
Note that inclusion of the log-population in $\mathbf{z}^{(\phi)}_{it}$ affects
susceptibility rather than infectivity, which is inverse to modelling
the force
of infection in the individual-based framework.
The influenza data yield an estimated coefficient of
$\hat{\beta}^{(\phi)}_{\mttt{log(pop)}} =
0.76$ (95\% CI: 0.50 to 1.01), which provides
strong evidence for such an agglomeration effect.
The variance of the random effect $b_i^{(\phi)}$ of the
spatio-temporal component is slightly reduced from
0.89 to
0.71,
reflecting a decrease in residual heterogeneity
between districts. The decay parameter is estimated to be slightly
smaller in
the extended model
[$\hat{d}=1.65$ (95\% CI: 1.45 to 1.86)]
and all other effects remain approximately unchanged (see Table~\ref
{tabfluEstimates}).
However, the predictive performance improves only minimally, for example,
the logarithmic score decreases from
0.5448
to 0.5447
($p = 0.66$).
This small change could be related to the random effects $b_i^{(\phi
)}$, which
replace parts of the population effect if it is not included as a covariate.
Indeed, there is correlation
($r_{\mathrm{Pearson}}=0.41$)
between log(pop$_i$) and $b_i^{(\phi)}$ in the model without an explicit
population effect in $\phi_{it}$ 
[see the scatterplot in \citet{supplementB}, Figure~5].

\subsection{Long-term forecast of the 2008 influenza wave}\label{secresultsflusims}

For further evaluation of the power-law models described in
Section~\ref{secresultshhh4}, we carry out a long-term
forecast of the wave of influenza in 2008. Specifically,
we simulate the evolution of the epidemic during the first 20 weeks in
2008 for
each model trained by the previous years and initialised by the
18 cases of the last week of 2007
(see the animation in \hyperref[supplemA]{Supplement~A}, for their spatial distribution).
Predictive performance is then evaluated by the final size
distributions and
by proper scoring rules assessing the empirical distributions induced by
the simulated counts both in the temporal and spatial domains.
Since the logarithmic score is infinite in the case of zero predictive
probability for the observed count, we instead use the
\citet{dawidsebastiani1999} score
\[
\operatorname{DSS}(P,y) = \frac{(y-\mu_P)^2}{\sigma_P^2} + \log \sigma_P^2,
\]
where $\mu_P$ and $\sigma_P^2$ denote the mean and the variance of $P$
[see also \citet{GneitingRaftery2007}].

%
\begin{figure}[b]
\begin{tabular}{@{}c@{\quad}c@{}}

\includegraphics{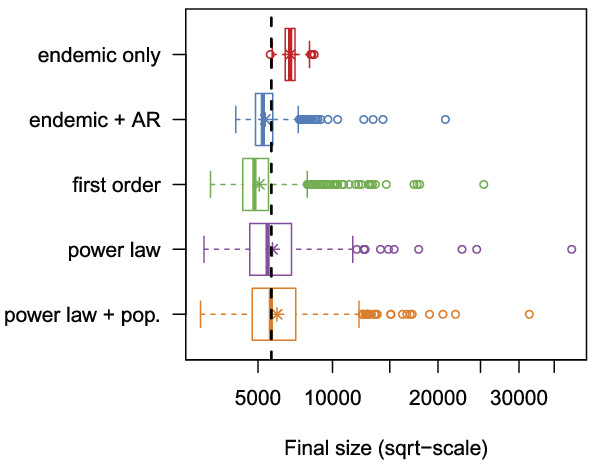}
 & \includegraphics{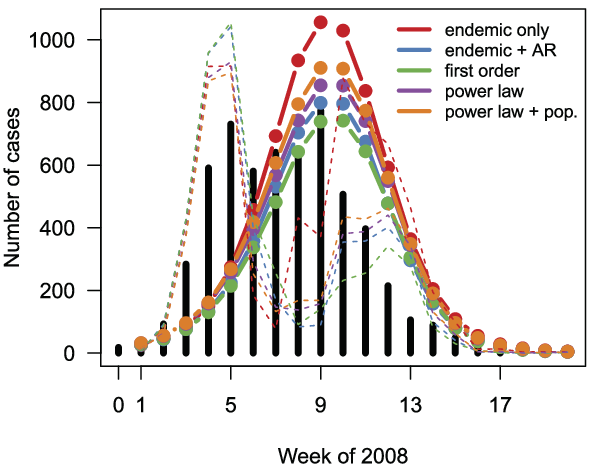}\\
\footnotesize{(a) Final size distributions (\raisebox{0.5ex}{$ \sqrt{}$}-scale). The}  & \footnotesize{(b) Time series of observed (bars) and mean}\\[-1.5pt]
\footnotesize{star in each box represents the mean, and}                                         & \footnotesize{simulated (dots) counts aggregated over all}\\[-1.5pt]
\footnotesize{the vertical dashed line marks the}                                           & \footnotesize{districts. Week 0 corresponds to the initial}\\[-1.5pt]
\footnotesize{observed final size of 5781 cases}                                            & \footnotesize{condition (2007-W52). The dashed lines show}\\[-1.5pt]
                                                                                            & \footnotesize{the (scaled) RPS (see also Table~\protect\ref{tabflusimsscores})}
\end{tabular}
\caption{Summary statistics of 1000 simulations of the wave of influenza
during the first 20 weeks of 2008 for five
competing models.}\label{figflusimsfinalsize}\label{figflusimstime}
\end{figure}

%
\begin{table}
\tabcolsep=5pt
\caption{Long-term predictive performance of 5 competing
models in the temporal and spatial dimensions measured by mean DSS and
RPS for the 2008 wave of influenza}\label{tabflusimsscores}
\begin{tabular*}{\tablewidth}{@{\extracolsep{\fill}}@{}lcccccc@{}}
\hline
\textbf{Model} & \multicolumn{2}{c}{\textbf{Time}} & \multicolumn{2}{c}{\textbf{Space}} & \multicolumn{2}{c@{}}{\textbf{Space--time}} \\[-6pt]
& \multicolumn{2}{c}{\hrulefill} & \multicolumn{2}{c}{\hrulefill} & \multicolumn{2}{c@{}}{\hrulefill} \\
& \textbf{DSS} & \textbf{RPS} & \textbf{DSS} & \textbf{RPS} & \textbf{DSS} & \textbf{RPS} \\
\hline
Endemic only & 27.03 & 149.77 & 7.85 & 15.39 & 2.91 & 1.31 \\
Endemic${}+{}$autoregressive & 31.36 & 112.15 & 7.59 & 15.04 & 2.58 & 1.26\\
First order & 26.46 & 108.61 & 7.51 & 15.63 & 2.50 & 1.26 \\
Power law & 16.41 & 110.20 & 7.36 & 14.75 & 2.29 & 1.25 \\
Power law${}+{}$population & 15.49 & 111.86 & 7.24 & 14.30 & 2.29 & 1.24 \\
\hline
\end{tabular*}
\end{table}

Figure~\ref{figflusimsfinalsize}(a) shows the final size
distributions of the
simulated waves of influenza during the first 20 weeks of 2008. Note
that model complexity increases from top to bottom and that we also
considered the naive endemic model, that is, independent counts, and the
model without a spatio-temporal component as additional benchmarks.
The endemic-only model, which decomposes disease incidence into
spatial variation across districts, a seasonal and a log-linear time trend,
overestimates the reported size of
5781 cases. It also does not allow
for much variability in the size of the outbreak as opposed to the
models with epidemic potential. The power-law models show the greatest
amount of variation but best meet the reported final size: the
power-law model without the population effect yields a simulated mean
of~6022 (95\% CI: 3126 to 10,808). The huge uncertainty seems plausible
with regard to the long forecast horizon
over a whole epidemic wave.

\begin{figure}

\includegraphics{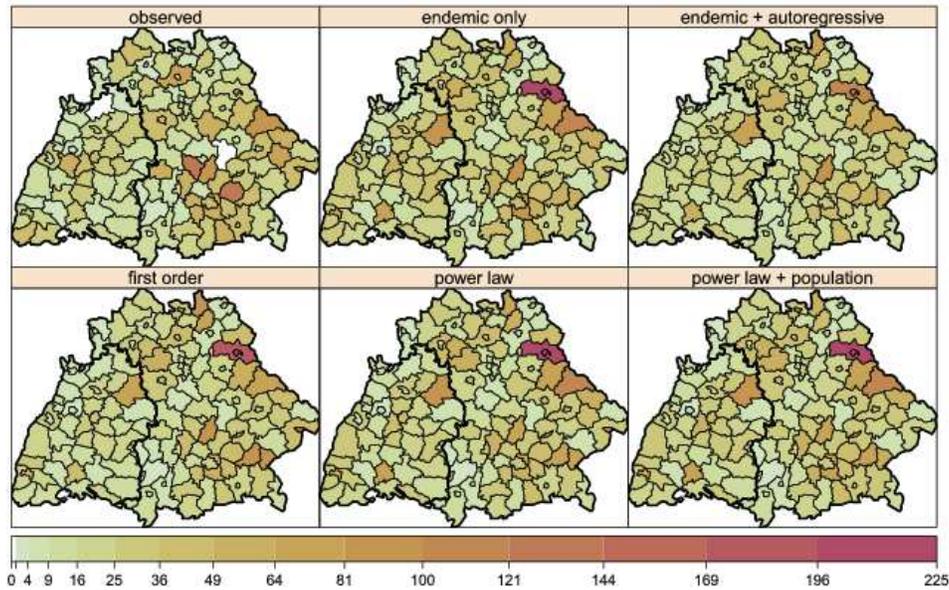}

\caption{Observed and mean simulated incidence (cases per 100 000 inhabitants)
aggregated over the 20 weeks forecast horizon (see
Figure~6 of Supplement~B for scatterplots).}\label{figflusimsspace}
\end{figure}

Figure~\ref{figflusimstime}(b) shows the time series of observed and mean
simulated counts aggregated over all districts.
In 2008, the wave grew two or more weeks earlier than in previous years trained
by the sinusoidal terms in the three components. This phenomenon cannot be
captured by the simulations, which are solely based on the observed pattern
during 2001--2007 and the distribution of the cases from the last week
of 2007.
Furthermore, instead of two peaks
as observed specifically in 2008, the simulations yield a single,
larger peak
where the power-law models on average induce the best amplitudes with
respect to
final size. The simulated spatial distribution of the cases
(see Figure~\ref{figflusimsspace}) is very similar among the various
models and
agrees quite well with the observed pattern.
Animations of the observed and mean simulated epidemics provide more insight
about the epidemic spread and are available in \hyperref[supplemA]{Supplement~A}.
It is difficult to see a clear-cut traveling-wave of influenza in the reported
data, which suggests that both an endemic component capturing immigration
as well as scale-free jumps via the spatio-temporal component, that is,
power-law
weights~$w_{ji}$, are important.
\hyperref[supplemA]{Supplement~A} also includes an animated series of weekly
probability integral
transform (PIT) histograms [\citet
{GneitingBalabdaouiRaftery2007}] using the
nonrandomised version for count data proposed by \citet
{czado-etal-2009}. These
sequential PIT histograms mainly reflect the above time shift of the
predictions.
More clearly than the plots, the mean scores in Table~\ref{tabflusimsscores} show
that predictive performance generally improves with increasing model complexity
and use of a power-law decay.

%

\section{Discussion} \label{secdiscussion}

Motivated by the finding of \citet{brockmannetal2006} that
short-time human
travel roughly follows a power law with respect to distance, we
investigated a power-law decay of spatial dependence between infections
in two modelling frameworks for spatio-temporal surveillance data.
A spatio-temporal point process model was applied to case reports of invasive
\mbox{meningococcal} disease,
and a multivariate time-series model was applied to counts of influenza
aggregated by week and district.
Since human mobility is an important driver of epidemic spread, the aim
was to
improve the predictive performance of these models
using a power-law transmission kernel with respect to
distance or neighbourhood order, respectively, where the decay is
estimated jointly with all other model parameters.

In both applications considered, the power-law formulations performed
better than previously used naive Gaussian or first-order interaction
models, respectively. Furthermore, alternative piecewise constant, but
otherwise unrestricted interaction models were in line with the
estimated power laws. This confirms that the power-law distribution
of short-time human travel translates to the modelling of infectious
disease spread. We note that the qualitative interaction models could
be replaced by (cubic) smoothing spline formulations, either in a continuous
[\citet{Schimek2000Eubank}] or in discrete fashion
[\citet{Schimek2000FahrmeirKnorr-Held}]. In order to penalise
deviations from
the power law, this should be done on a log--log scale, where the power
law is a simple linear relationship. However, data-driven estimation
of the smoothing parameter may become difficult.

The heavy tail of the power law allows for long-range dependence
between cases, which accordingly increased the importance of the
epidemic component in both models. An alternative formulation of
spatial interaction with occasional long-range transmission was used
by \citet{diggle2006}, who added a small distance-independent
value to
a powered exponential term of the scaled distance.
However, this offset and the power parameter are poorly identified.
For the 2001 UK foot-and-mouth disease epidemic,
\citet{keelingetal2001} observed a power-law-like, sharply
peaked transmission
kernel, and \citet{chissterferguson2007} subsequently found that
the power law
(\ref{eqnsiafpowerlaw}) yields a much better fit than
the offset kernel or other functional forms, which is in accordance
with our
results for the spread of human infectious diseases.

Regions at the edge of the observation window are missing
potential sources of infection from the unobserved side of the border.
To capture unobserved heterogeneity due to
immigration/edge effects, the count data model includes region-specific random
effects $b_i^{(\nu)}$ in the endemic component.
However, there was no clear pattern in their estimates 
with respect to regions being close to the border or not [\citet
{supplementB}, Figure~4].
In contrast, the IMD data supported edge effects, specifically
concerning the
border to the Netherlands. The spatial occurrence of cases met our
simplistic approach of including the distance to the border as a
covariate in the
endemic component. This ignores that immigration might be more
important in large metropolitan areas attracting people from abroad regardless
of the location within Germany.
A better way of accounting for edge effects would thus be to explicitly
incorporate immigration data. For instance,
\citet{geilhufeetal2012} used incoming road or air traffic from
outside North
Norway as a proxy for the risk of importing cases of influenza, which
led to
improved predictive performance while also accounting for population in the
spatio-temporal component.

Scaling regional susceptibility by population size proved very
informative also
for influenza in Southern Germany: more populated regions seem to
attract more
infections from neighbours than smaller regions, which reflects commuter-type
imports [see \citet{viboudetal2006}, and
\citet{KeelingRohani2008}, Section~6.3.3.1].
An exception of such a population effect in the spatio-temporal
component might
be seasonal accumulations in low-populated touristic regions.
In the point process model for the IMD cases, the effect of population density
on infectivity was less evident, which might be related to the very
limited size
of the point pattern with less than 100 cases per year over all of Germany.

Another limitation of the IMD data set is tied locations of cases due to
censoring at the postcode level. For the power law to be identifiable, we
randomly sampled the locations from discs of radius 0.59 km
around the centroid of the respective postcode area, and verified that our
results are insensitive to the random seed. Note that choosing a larger
radius of, for example, 3 km, leads to less pronounced weight towards
zero distance but
yields otherwise similar results, especially concerning the
relative performance of the various interaction functions.

We considered power laws as a description of spatial dispersal of
infectious diseases as motivated by human travelling behaviour.
Concerning temporal dispersal, power laws are usually not an appropriate
description of the evolution of infectivity over time. Infectious diseases
typically feature a very limited period of infectivity after the incubation
period, since an infected individual will receive treatment and typically
restrict its interaction radius upon the appearance of symptoms.
Due to the small number of cases in the IMD data, we could not estimate
a parametric temporal interaction function $g(t)$ and simply assumed constant
infectivity during 30 days as in \citet{meyeretal2011}.
More generally, $g(t)$ could represent an increasing
level of infectivity beginning from exposure, followed by a plateau and then
decreasing and eventually vanishing infectivity
[\citet{lawsonleimich2000}, Section~5.3].
In the multivariate time-series model, the counts were restricted
to only explicitly depend on the previous week. This is reasonable if the
generation time, the time consumed by an infective to cause a secondary case,
is not larger than the aggregation time in the surveillance data.
For human influenza, \citet{cowlingetal2009} report a mean
generation time of
3.6 days (95\% CI: 2.9 days to 4.3 days).

Long-term simulated forecast of the 2008 influenza wave
confirmed that the power-law model yields better predictions.
However, the model was not able to describe the onset in 2008,
which was two weeks earlier than in the years 2001--2007. For this
to work, it would be necessary to further enrich the model by
external processes such as vaccination coverage [as in \citet
{herzogetal2011}]
or climate conditions [\citet{fuhrmann2010,willemetal2012}]
entering as covariates in the endemic and/or epidemic components.
An alternative approach has been used by \citet{fanshaweetal2008},
where seasonality parameters were allowed to change from year to year
according to a random walk model. Implementation would then
require Markov chain Monte Carlo or other more demanding techniques for
inference.
Despite the open issue of dynamic seasonality, the simulated final size and
spatial distribution matched the reported epidemic quite well.

This success also suggests that under-reporting of influenza was
roughly constant over time. For instance, the
4 districts which did not report any cases
during the 2008 forecast period (SK Kempten, SK Memmingen, LK Kelheim,
and SK Aschaffenburg)
only reported
1, 0, 20, and 4 cases in
total during 2001--2007.
However, we can only model the effectively reported number of
cases, which may be affected by time-varying attention drawn to
influenza in the media. 
Syndromic surveillance systems aim to unify
various routinely collected data sources, for example, web searches for
outbreak detection and monitoring
[\citet{josseranetal2006,hulthetal2009}], and may thereby
provide a
more realistic picture of influenza.

Prospective detection of outbreaks is also possible based on the
count data model presented here. A statistic could be based on
quantiles of the distribution of $Y_{i,t+1} \vert\mathbf{Y}_{\cdot,t}$,
for example, an
alarm could be triggered if the actual observed counts at $t+1$ are
above the
99\% quantile, say, [\citet{heldetal2006a}]. Note that by including
seasonality in the model, a yearly wave at the beginning of the year
would be
`planned' and not necessarily considered a deviation from default behaviour.

Our power-law approach is very useful in the absence of movement
network data
(e.g., plane and train traffic).
However, if such data were available [\citet{lazeretal2009}],
neighbourhood
weights $w_{ji}$ in the count data model could instead be based on the
connectivity between regions, 
which was investigated
by \citet{schroedleetal2011a} for the spread of Coxiellosis in
Swiss cows
and by \citet{geilhufeetal2012} for the spread of influenza in
Northern Norway.
In recent work, \citet{brockmannhelbing2013} introduce the
`effective distance' to describe the 2009 H1N1
influenza pandemic. Their approach relates to what has already been termed
`functional distance' by \citet{brownhorton1970}, that is,
a function of (inter-)regional properties like population and commuter or
travel flows such that it ``reflects the net effect of entity properties upon
the propensity of the entities to interact'' [\citet{brownholmes1971}].
A recent example of using telephone call data as a measure of human interaction
can be found in \citet{rattietal2010}. 
Another fruitful area of future research is the statistical analysis of
age-stratified surveillance data. Contact patterns vary across age
[\citet{mossongetal2008,truscottetal2012}], calling for a
unified analysis across
age groups and regions. 


\begin{appendix}
\section*{Appendix: Software} \label{appendixsoftware}

All calculations have been carried out in the statistical software environment
\textsf{R} 3.0.2 [\citet{Rbase}].
Both model frameworks and their power-law extensions presented in this
paper are implemented in the \textsf{R} package \pkg{surveillance}
[\citet{Rsurveillance}] as of version 1.6-0 available from the
Comprehensive
\textsf{R} Archive Network (\href{http://www.CRAN.R-project.org}{CRAN.R-project.org}).
The two analysed data sets are included therein as
\texttt{data("imdepi")} (courtesy of the German Reference Centre for
Meningococci)
and \texttt{data("fluBYBW")} [raw data obtained from the German national
surveillance system operated by the \citet{RKIsurvstat}].
The point process model (\ref{eqntwinstim}) for individual point-referenced
data can be fitted by the function \texttt{twinstim()}, and
the multivariate time-series model (\ref{eqnmeanHHH}) for count data
is estimated by \texttt{hhh4()}.
The implementations are flexible enough to allow for other specifications
of the spatial interaction function $f$ and the weights $w_{ji}$, respectively.
A related two-component epidemic model [\citet{hoehle2009}],
which is designed
for time-continuous individual surveillance data of a closed population
with a
fixed set of locations, for example, for farm- or household-based
epidemics, is also
included as function \texttt{twinSIR()}.
The application of all three model frameworks in \textsf{R} is
described in detail in \citet{meyeretal2014}.

Spatial integrals in the point process likelihood have been evaluated using
cubature methods implemented in the \textsf{R} package \pkg{polyCub}
0.4-3 [\citet{RpolyCub}].
Maps have been produced using \pkg{sp}
1.0-15 [\citet{Bivandetal2013}]
and animations using \pkg{animation}
2.2 [\citet{xie2013}].
\end{appendix}

\section*{Acknowledgements}
This work was presented at the \emph{Summer School on Topics in Space--Time
Modeling and Inference} 
at Aalborg University, May 2013, which enabled fruitful discussions
with its
participants.
These also gave rise to the efficient cubature rule for isotropic
functions over polygonal domains elaborated in \citeauthor{supplementB} [(\citeyear{supplementB}), Section~2.4]
with valuable support by Emil Hedevang and Christian Reiher.
We thank Michaela Paul for technical support on the original count data
model, as well as Johannes Elias and Ulrich Vogel from the German Reference
Centre for Meningococci for providing us with the IMD data.
We also appreciate helpful comments by Julia Meyer, Michael H\"ohle,
the Editor Tilmann Gneiting, and two anonymous referees.

\begin{supplement}\label{supplemA}
\textbf{Supplement A: Animations of the IMD and influenza epidemics\\}
(\url{http://www.biostat.uzh.ch/static/powerlaw/}).
\begin{itemize}
\item Observed evolution of the IMD and influenza epidemics.
\item Simulated counts from various models for the 2008 influenza wave.
\item Weekly mean PIT histograms for these predictions.
\end{itemize}\vspace*{12pt}

\sname{Supplement B}
\stitle{Inference details, integration of isotropic functions over
polygons, and additional figures and tables\\}
\slink[doi]{10.1214/14-AOAS743SUPPB} 
\sdatatype{.pdf}
\sfilename{aoas743\_suppB.pdf}
\sdescription{
\begin{itemize}
\item Details on likelihood inference for both models.
\item Integration of radially symmetric functions over polygonal domains.
\item Additional figures and tables of the power-law models for invasive
meningococcal disease and influenza.
\end{itemize}
}
\end{supplement}

%

\printaddresses
\end{document}